\documentclass[aps,twocolumn,prl,showpacs,color,psfig,epsf]{revtex4}
\usepackage{amsmath}
\usepackage{color}
\usepackage{amsfonts}
\usepackage{epsf}
\usepackage{graphicx}
\newcommand\Er{Er}

\begin{document}

\title{Colloids in cholesterics: Size-dependent defects and non-Stokesian microrheology}


\author{J. S. Lintuvuori$^1$, K. Stratford$^2$, M. E. Cates$^1$ and D. Marenduzzo$^1$}
\affiliation{$^1$SUPA, School of Physics and Astronomy, University of Edinburgh, Mayfield Road, Edinburgh, EH9 3JZ, UK;\\
$^2$EPCC, School of Physics and Astronomy, University of Edinburgh, Mayfield Road, Edinburgh, EH9 3JZ, UK.}

\begin{abstract}
We simulate a colloidal particle (radius $R$) in a cholesteric liquid crystal (pitch $p$) with tangential order parameter alignment at the particle surface. The local defect structure evolves from a dipolar pair of surface defects (boojums) at small $R/p$ to a pair of twisted disclination lines wrapping around the particle at larger values. 
On dragging the colloid with small velocity $v$ through the medium along the cholesteric helix axis (an active microrheology measurement) we find a hydrodynamic drag force that scales linearly with $v$ but superlinearly with $R$ -- in striking violation of Stokes' law, as generally used to interpret such measurements.
\end{abstract}

\maketitle

Understanding the properties of colloidal particles moving in complex fluids is a challenging goal of soft matter physics and fluid dynamics. The case of a Newtonian fluid is very well understood, and it is accurately described by Stokes' law~\cite{landau}, which states that the viscous force felt by a particle moving at a constant speed $v$ is linearly proportional to the solvent viscosity, $\eta$, to $v$ itself and to the particle radius $R$. This fundamental and elegant result is exploited routinely in active microrheology, a modern technique which consists of dragging a particle in a fluid, and measuring the force--velocity relation. From these data, and the knowledge of the probe size, one can obtain an estimate of a fluid viscosity (effective microviscosity)~\cite{Squires}, $\eta^{\mathrm{(active)}}_{\mathrm{micro}}=\frac{F}{6\pi vR}$, which is directly a property of the local environment. This is often preferrable to a bulk rheology experiment, e.g. when addressing the jamming of dense suspensions~\cite{L.G.Wilson}.  

What happens when a particle, instead, moves in fluid with broken symmetry? Here, much less is known. 
Sedimentation or ``falling ball'' experiments showed long ago that the drag force in nematic liquid crystals (in which molecular alignment is governed by a director field $\hat{ \bf n}$) is anisotropic, as one might expect for a fluid of spontaneously broken rotational symmetry. More recently, passive microrheology experiments and theories have quantified the ratio of the viscosities along and perpendicular to the local director field, and found these to differ by about a factor of 2~\cite{NematicColloidDrag}. These results were obtained in the linear regime where the Ericksen number \Er, quantifying the ratio between viscous and elastic effects, is small: the particle moves ``slowly''. A series of numerical simulations have also addressed the case of intermediate to high \Er, corresponding to nonlinear (in $v$) microrheology. It was found that the ratio between the viscosities becomes much smaller in this case, and the defect structure changes, e.g. from a Saturn ring to a dipole~\cite{Tanaka-and-DePablo,drag_colloidal_lc}.

Here we consider the hydrodynamics of a colloidal particle (radius $R$) inside a cholesteric liquid crystal, which differs from the nematic by having a helical twist to the director field with a pitch $p$. (This additional order breaks translational as well as rotational symmetry.) We focus on the case of tangential anchoring in which the director field at the colloid-fluid interface lies everywhere parallel to the surface of the particle. We show that the resulting defect structure depends crucially on the ratio $R/p$. On increasing this ratio we observe a crossover from a dipolar configuration, with two defect patches on opposite sides of the particle, to a twisted set of disclination lines of opposite chiralities wrapping the colloid. 

This crossover could profoundly influence the many-body self-assembly of colloids within the liquid crystal to form organized structures~\cite{nematic_self_assembly}: we defer this issue to future work. Here we address a much simpler problem that is nonetheless directly addressable in the laboratory. We
model an active microrheology experiment, in which we drag an isolated colloid through the liquid crystal, defining two effective viscosities for motion along and perpendicular to the cholesteric helix. We find that, in contrast to the nematic case, the ratio between these two effective viscosities, even in the small Er regime, depends strongly on the size of the probe. This is because the effective drag on a particle moving along the cholesteric helix is superlinear in its size $R$ (with exponent $\sim 1.7$), in striking violation of Stokes' law. This presents an instructive case in which microrheology delivers a probe-size dependent ``viscosity", essentially unrelated to any bulk macroscopic value, and only indirectly related to the material parameters of the medium even as defined on the length scale $R$.
Finally, we address the case of intermediate Ericksen number ($\mathrm{Er}\approx 1$), where the disclinations wrapping the dragged colloid are displaced downstream to form a double twisting disclination wake.

{\it Methodology:} The thermodynamics of the cholesteric solvent is determined by the Landau--de Gennes free energy $\cal{F}$. Its density ${f}$ is expressed in terms of a (traceless and symmetric) tensorial order parameter $\mathbf{Q}$~\cite{beris} as
\begin{align}
{f} & = \tfrac{A_0}{2} \bigl( 1 - \tfrac{\gamma}{3} \bigr) Q_{\alpha \beta}^2 
           - \tfrac{A_0 \gamma}{3} Q_{\alpha \beta}Q_{\beta \gamma}Q_{\gamma \alpha}
           + \tfrac {A_0 \gamma}{4} (Q_{\alpha \beta}^2)^2 \notag \\ 
	 & \quad + \tfrac{K}{2}\bigl( \nabla_{\beta}Q_{\alpha \beta}\bigr)^2
	   + \tfrac{K}{2} 
           \bigl( \epsilon_{\alpha \gamma \delta} \nabla_{\gamma} Q_{\delta \beta} 
           + 2q_0 Q_{\alpha \beta} \bigr)^2. 
\label{eq:FreeEnergy}
\end{align}
Here $A_0$ is a constant (setting the energy scale), $K$ is an elastic constant, $q_0=2\pi/p$, and $\gamma$ is a temperature-like control parameter governing proximity to the isotropic-to-cholesteric transition. In our notation Greek indices denote Cartesian components and summation over repeated indices is implied; $\epsilon_{\alpha\gamma\delta}$ is the permutation tensor.

We employ a 3D hybrid Lattice Boltzmann (LB) algorithm~\cite{LBLC} to solve the Beris-Edwards equations for the evolution of the $\mathbf{Q}$ tensor~\cite{beris}
\begin{equation}
D_t \mathbf{Q} 
= \Gamma  \Bigl( \tfrac{-\delta {\cal F}}{\delta \mathbf{Q}} + \tfrac{1}{3}\, 
\text{tr} \Bigl( \tfrac{\delta {\cal F}}{\delta \mathbf{Q}} \Bigr) \mathbf{I} \Bigr)  .
\label{eqQevol}
\end{equation} 
Here $\Gamma$ is a collective rotational diffusion constant and $D_t$ is the material derivative for rod-like molecules~\cite{beris}. The term in brackets is the molecular field, $\mathbf{H}$, which ensures that in the absence of flow $\mathbf{Q}$ evolves towards a minimum of the free energy. The velocity field obeys the continuity equation and a Navier-Stokes equation with a stress tensor, $\Pi_{\alpha\beta}$, generalised to describe liquid crystal hydrodynamics~\cite{LBLC}.
For the hydrodynamics, the colloid is represented by the standard method of bounce-back on links~\cite{ladd}, where the LB distributions which link solid to fluid nodes in the underlying discretised space are used to impose the appropriate boundary conditions at each step. Tangential boundary conditions for $\hat{\bf n}$ are imposed on the particle surface ($\hat{\bf n}$ is free to rotate in the tangent plane). Order parameter variations create an additional elastic force acting on the particle, $F^{\rm el}$, which we computed by integrating the stress tensor $\Pi_{\alpha\beta}$ over the particle surface~\cite{zumer_colloid},
$F_{\alpha}^{\rm el}=\int dS \Pi_{\alpha\beta} \hat{\nu}_{\beta},$ 
where $\hat{\nu}_{\beta}$ is the local normal to the colloid surface. Furthermore, the particle may rotate due to elastic and hydrodynamic torques.

The thermodynamics of chiral liquid crystals is determined by the chirality $\kappa$ and the reduced temperature $\tau$, which are given in terms of previously defined quantities as:
$\kappa = \sqrt{{108 K q_0^2}/{A_0 \gamma}}$ and 
$\tau = 27 (1-\gamma/3)/\gamma.$ 
The physics of a colloidal particle moving in the liquid crystal is controlled by the Ericksen number
\begin{equation}
{\mathrm{Er}}  = {\gamma_1 vR}/{K},~\mathrm{with}~\gamma_1={2q^2}/{\Gamma}
\end{equation}
which measures the ratio between viscous and elastic forces.  In the expression for Er, $\gamma_1$ is the rotational viscosity of the liquid crystal, $q$ is the degree of ordering in the system (for the uniaxial case with director $\hat{\bf n}$, $Q_{\alpha\beta}=q(\hat{n}_\alpha \hat{n}_\beta-\delta_{\alpha\beta}/3)$). The Ericksen number should determine which dynamical regime we are in. As we shall see, the presence of the cholesteric helix brings another control parameter into play: $R/p$. This changes the physics completely with respect to the nematic case.

We give most of our results in simulation units~\cite{simulation-units,LBLC}. To convert them into physical ones, we can assume an elastic constant of 6.5 pN, and a rotational viscosity of 1 poise. (These values hold for typical materials, and a colloidal diameter of 1 $\mu$m.) In this way, one can show that the simulation units for force, time and velocity map onto 90 pN, 1 $\mu$s, and 0.03$\mu$m/s respectively.

{\it Results:} We first discuss the disclination structure around the colloidal particle. In the case of planar anchoring, a colloid in a nematic host leads to two surface defects of topological charge $+1$ (boojums)~\cite{boojums}. In the case of a chiral nematic host, we find the behavior is determined by the ratio between the colloidal size and the cholesteric pitch. For $R/p$ significantly smaller than $1/2$, we observe two defect patches on opposite sides of the particles (Fig.~1a). On the other hand, if the particle size exceeds the pitch, then the minimum energy configuration becomes a twisted pair of disclination lines of opposite chirality, each of which wraps around the particle (Fig.~1b and 1c). These can be viewed as elongated boojums, in which a 3d disclination line of charge +1/2 lies adjacent to the surface terminating in a 2d defect of a charge +1/2 on the surface at each end. The length of these ``chiral rings'' is not constant, rather it increases with the size of the particle, and for very large colloids ($R = p$) further disclination rings appear at the top and bottom of the particle, leading to a more uniform coverage of the surface. 
Unlike the transition between Saturn rings and dipoles in standard nematic colloids \cite{Fukuda}, this conformation change does not depend on the degree of surface anchoring. It should thus be more easily observable experimentally, purely by changing the particle size $R$ or the chiral fraction (and hence $p$) in a mixed nematogen. Just as for the nematic case, the two defect patterns which we found have sufficiently different geometry that they should mediate quite distinct two-body and many-body effective interactions, creating new avenues for directing the self-assembly of colloids by tuning properties of the surrounding matrix \cite{nematic_self_assembly}.

\begin{figure}
\label{disclination-structure}
\includegraphics[width=8.cm]{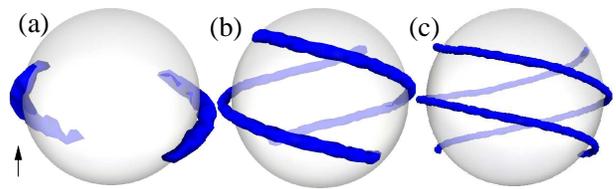}
\caption{Defect structure close to the particle. The arrow denotes the direction of the cholesteric helix. For a small particle (a), $R/p=1/4$, there are two defect patches on opposite sides. These elongate to form twisted disclination pairs for particles with $R/p\ge 1/2$: (b) $R/p=1/2$ and (c) $R/p=3/4$.}
\end{figure}

Having seen the importance of the control parameter $R/p$ in determining the statics of colloids in cholesterics, it is natural to ask if this length scale ratio leads to different physics (from the nematic case, $R/p = 0$) for their hydrodynamics as well. To address this question, we studied the dynamics of colloids of various radii in response to a constant pulling force, which is either parallel or perpendicular to the axis of the cholesteric helix. First we consider the $\mathrm{Er}\ll 1$ regime.

Fig.~2 shows the average velocity--force relations observed, for dragging perpendicular and parallel to the cholesteric helix, at various $R/p$ ratios. We fitted the data by means of the following formula for the drag force:
\begin{equation}\label{perpendicular-drag}
f(v) = f_y+a(R)vR
\end{equation}
where $f_y$ and $a(R)$, with dimensions of force and viscosity respectively, are parameters of the fit. We allow for a finite ``yield force'' $f_y$, which can be seen as the microrheological analogue of a yield stress. 
\begin{figure}
\label{velocity-force}
\includegraphics[width=\columnwidth]{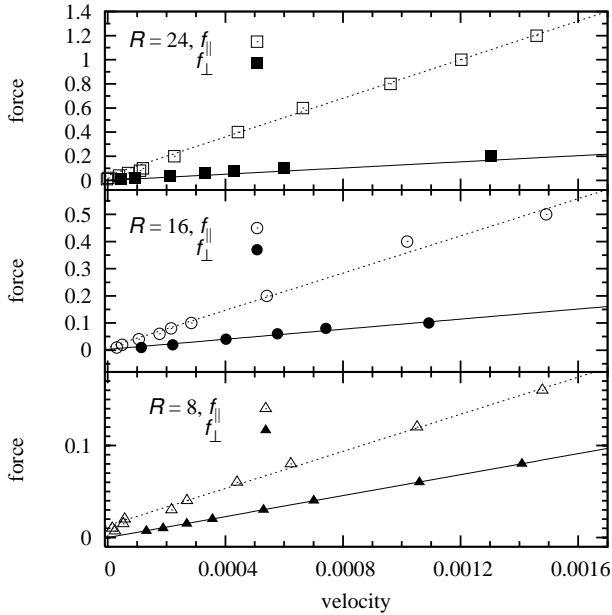}
\caption{(a) Average velocity--force characteristics for particles with $R/p$ equal to 3/4, 1/2 and 1/4 respectively. The fits are with the formulas shown in the text. The averages were calculated from the last $2.5-5\times 10^4$ simulation steps. At steady state, the velocity along the helix shows slight periodic fluctuations around the average value.}
\end{figure}
Our data show that the yield force, while possibly non-zero, is very small~\cite{note_parameter_table}, so that it does not appreciably change our analysis~\cite{note_yield}. Much more striking than any yield force is the extended regime in which the force is linear in velocity. (This would be the usual criterion for accepting an active microrheology measurement as measuring a length-scale dependent, but nonetheless linear, effective viscosity.) Fig.~3 shows the dependence of $f/v\sim aR$ on $R$, for dragging both perpendicular and parallel to the cholesteric helix. We fitted these data with a power law, $f/v\sim R^{\alpha}$. The data corresponding to dragging perpendicular to the helical axis are comparable with those reported for nematic colloids: $a(R)\sim R^{\alpha-1}$ is almost constant in agreement with Stokes' law~\cite{note_scaling}. 
The fit gives $\alpha \approx 1.07 \pm 0.10$; (note that the prefactor can depend on the in-plane angle between force and director). In contrast, pulling along the helix shows very different physics, with $\alpha \approx {1.7}$. 
\begin{figure}
\label{force-per-velocity}
\includegraphics[width=\columnwidth]{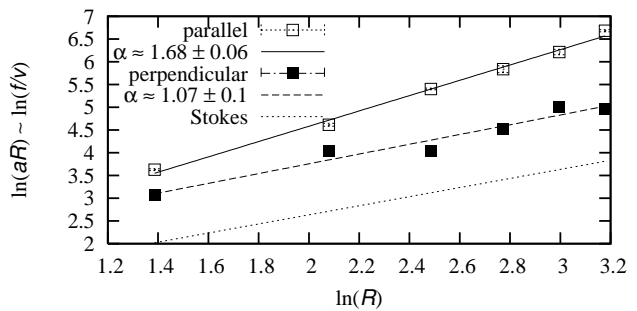}
\caption{$\ln(aR)$--$\ln(R)$ relation for 
colloids dragged perpendicular (closed squares) and along (open squares) the helix. Dashed and solid lines are linear fits of $\ln(aR)=\beta+\alpha\ln(R)$. The dotted line is Stokes' law $f/v=6\pi\eta R$, for $\eta=0.1$.}
\end{figure}

One way to discuss these results is to insist on Stokes' law as defining an effective microviscosity $\eta(R) = a(R)/6\pi$, in which case for motion along the pitch $\eta(R) \sim R^{0.7}$. This strong size effect could be attributable to the $R$ dependence of the defect structure or to non-Stokes features of the flow. But unless these effects are fully understood, active microrheology is uninformative for chiral nematics, in the sense that it measures a combined property of the probe and environment, rather than any material property (local or otherwise) of the medium alone. Remarkably, this breakdown arises at very low \Er, in a regime where the $v$ dependence is still linear (Fig.~2). Our results can be illuminated by considering the bulk shear rheology of cholesterics undergoing permeation flow along the pitch direction. In this case, in any finite sample there is a regime of linear rheology, but the shear viscosity increases linearly with system size, diverging in the thermodynamic limit~\cite{permeation,chol-shear-experiment}. Crudely treating the colloid radius $R$ as an effective sample size then gives $\alpha = 2$, not far from the observed value ($\alpha\approx 1.7$).

To explore further the physics involved, Fig.~4 show director profiles for a stationary colloid, and ones moving along the helical axis with Er $\approx 0.03$, and Er $\approx 0.7$. In the stationary case the liquid crystal accommodates the colloid by  a local deformation involving the disclination pattern of Fig.~1c. Interestingly, even at very small forcing the cholesteric layers slightly bend throughout the whole simulation box: the elastic distortions induced by the colloid motion appear long ranged (Fig.~4b). Here, the disclination rotates around the helical axis and is slightly displaced in the downstream direction, but still remains on the particle surface. At larger \Er, (Fig.~4c), the bending of layers is more pronounced, and the disclination is fully displaced downstream to form a double twisting disclination ``wake'' (Fig.~4d). 
\begin{figure}
\label{director-snapshots}
\includegraphics[width=0.9\columnwidth]{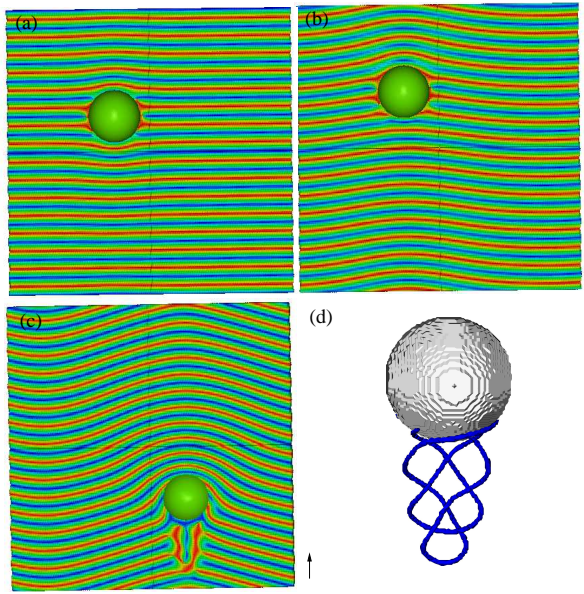}
\caption{Snapshots of a colloid with $R/p = 3/4$ in a cholesteric: (a) stationary; (b) moving along the helical axis ($\mathrm{Er}\approx 0.03$) and (c) likewise with $\mathrm{Er}\approx 0.7$. (d) Disclination structure corresponding to (c). (a)-(c) were produced with QMGA \cite{qmga}. Colors denote the magnitude of $\hat{n}_x$ (the helix is along $\hat{z}$).} 
\end{figure}

In summary, we have reported lattice Boltzmann simulations of a colloidal particle embedded in a cholesteric liquid crystal. Focussing on tangential anchoring of the liquid crystal at the colloidal surface, we have shown that by changing the ratio between particle size and cholesteric pitch it is possible to control the topology of the local defects imposed by the presence of the colloid. For small particles, the equilibrium configuration has two defects at opposite poles of the sphere. Increasing the particle size leads to a texture with two twisted disclination lines of opposite polarity, wrapping around the colloid. Both these novel defect structures should be detectable in experiments with crossed polarizers. 
When pulling the particle along the cholesteric helix, we have found the drag force is linear in $v$ but depends superlinearly on particle size, in violation of Stokes' law which is the usual basis for interpreting active microrheology experiments~\cite{Squires}. Laser tweezers experiments (similar to those in nematic~\cite{tweezers_nem} and in twisted nematic cells~\cite{tweezers_twist}) should be able to test our predicted superlinear scaling. 

We hope that our results will stimulate further experiments~\cite{3Dmanipulation} and theoretical work on the microrheology of colloids in cholesterics. Furthermore,  similar phenomenology could be expected in  various complex fluid phases characterised by a spatially variable order parameter, e.g. smectics, blue phases, and lyotropic cubic liquid crystals, where one expects permeation flow to occur.

Acknowledgements: Work funded in part by EPSRC Grants EP/E030173 and EP/E045316. MEC is funded by the Royal Society.


\begin{thebibliography}{99}
\bibitem{landau} L.~D.~Landau and E.~M.~Lifshitz, {\it Fluid Mechanics}, 
Pergamon Press, Oxford (1987).
\bibitem{Squires} T.~M.~Squires and T.~G.~Mason, {\it Ann. Rev. Fluid Mech.} {\bf 42}, 413 (2010).
\bibitem{L.G.Wilson}
L.~G.~Wilson {\em et al.},
{\it J. Phys. Chem. B} {\bf 113}, 3806 (2009). 
\bibitem{NematicColloidDrag}
R.~W.~Ruhwandl and E.~M.~Terentjev, {\it Phys. Rev. E} {\bf 54}, 5204 (1996);
H.~Stark and D.~Ventzki, {\it Phys. Rev. E} {\bf 64}, 031711 (2001);
J.~C.~Loudet, P.~Hanusse and P.~Poulin {\it Science} {\bf 306}, 1525 (2004).
\bibitem{Tanaka-and-DePablo}
T.~Araki and H.~Tanaka, {\it J. Phys. Condens. Matt.}
{\bf 18}, L193 (2006);
B.~T.~Gettelfinger {\em et al.},
{\it Soft Matter} {\bf 6}, 896 (2010).
\bibitem{drag_colloidal_lc} A.~A.~Verhoeff {\em et al.}, 
{\it Soft Matter} {\bf 4}, 1602 (2008).
\bibitem{nematic_self_assembly} J.~C.~Loudet, {\em et al.}, {\it Langmuir} {\bf 20}, 11336 (2004).
\bibitem{beris} 
A.~N.~Beris and B.~J.~Edwards, {\it Thermodynamics of Flowing Systems},
Oxford University Press, Oxford, (1994).
\bibitem{LBLC}
D.~Marenduzzo et al., {\it Phys. Rev. E} {\bf 76}, 031921 (2007); M.~E.~Cates et al., {\it Soft Matter} {\bf 5}, 3791 (2009).
\bibitem{simulation-units} Parameters were (in simulation units): 
$A_0=1.0,~K \simeq 0.065,~\xi=0.7,~\gamma = 3.0$, $p=32$, $q=1/2$ and $\Gamma =0.5$. They give $\tau=0$, $\kappa=0.3$ and $\gamma_1=1$. Three periodic 
cubic simulation box sizes ($64^3,~128^3,~256^3$) were used.
\bibitem{ladd} N.-Q. Nguyen and A. J. C. Ladd, {\it Phys. Rev. E}
{\bf 66}, 046708 (2002).
\bibitem{zumer_colloid}M.~Skarabot {\em et al.} 
{\it Phys. Rev. E} {\bf 77}, 061706 (2008);  M.~Conradi {\em et al.},  
{\it Soft Matter} {\bf 5}, 3905 (2009).
\bibitem{boojums} P. Poulin and D. A. Weitz {\it Phys. Rev. E} {\bf 57}, 626 (1998).
\bibitem{Fukuda} J. Fukuda {\em et al.}, 
{\it Eur. Phys. J. E} {\bf 13}, 87 (2004).
\bibitem{note_parameter_table}See supplementary material at for a table of the obtained parameters.
\if{ 
Parameters obtained from fits of equation (\ref{perpendicular-drag}) were as follows. For dragging perpendicular to the helix, $f_y$ and $a$ were respectively 
$-0.0003\pm 0.0001$ and $5.34\pm0.03$ ($R=4$); 
$-0.0002\pm 0.0002$ and $7.12\pm0.02$ ($R=8$); $0.004\pm 0.002$ and  
$4.7\pm0.2$ ($R=12$); $0.003\pm 0.006$ and $5.8\pm 0.6$ ($R=16$); 
$-0.005\pm 0.002$ and $7.5\pm 0.2$ ($R=20$); $0.013\pm 0.003$ and  
$6.0\pm 0.2$ ($R=24$).
For dragging along the helix, $f_y$ and $a$ were respectively 
$0.0024\pm 0.001$ and $9.4\pm0.2$ ($R=4$);  $0.013\pm 0.003$ and  $12.6\pm0.4$
($R=8$); $-0.011\pm 0.003$ and $18.5\pm0.3$ ($R=12$); $0.01\pm 0.02$ and
 $21.4\pm1.6$ ($R=16$); $0.03\pm 0.03$ and $25.0\pm1.5$ ($R=20$); $0.04\pm 0.02$ and  $33.3\pm 0.8$ ($R=24$).
}\fi
\bibitem{note_yield} Experiments on sheared cholesterics showed that flow along the helix does lead to a linear regime and no yield stress~\cite{chol-shear-experiment}, which is also predicted by theory~\cite{permeation}. In the microrheological case, the colloid induces a small stress at zero flow, but this vanishes when integrated over its surface. Very small forces drive an oscillatory motion -- whether there is a yield force or not depends on whether the particle motion over a long time averages out to 0 or exhibits a drift. Our simulations favour the latter. 
\bibitem{note_scaling} Stokes' law, when applicable, describes the limit of zero volume fraction. Our studies use periodic boundary conditions and therefore effectively entail a finite volume fraction $\phi = 4\pi R^3/3L^3$ where $L$ is the box size. For particle size $R = 8$ we have investigated this $\phi$ dependence and find that a 64-fold change in $\phi$ alters the $f$ values by no more than 30\% (which is consistent with the finite size corrections of Hasimoto~\cite{hasimoto} for a simple fluid). In Fig.3, the $\phi$ values for different $R$ vary only 16-fold, from $1.0 \times 10^{-3}$ and $1.6 \times 10^{-2}$; moreover, this variation is not correlated with $R$. Therefore the reported $R$ scalings cannot be directly attributed to finite $\phi$ effects, although these could contribute to the error in fitted exponents.
\bibitem{chol-shear-experiment} M.~Yada, J.~Yamamoto and H.~Yokoyama {\it Langmuir} {\bf 19}, 3650 (2003).
\bibitem{permeation} W.~Helfrich {\it Phys. Rev. Lett.} {\bf 23}, 372 (1969);
T.~C.~Lubensky {\it Phys. Rev. A.} {\bf 6}, 452 (1972);
A.~D.~Rey {\it Phys. Rev. E} {\bf 65}, 022701 (2002); {\it J. Rheol.} {\bf 46}, 225 (2002).
\bibitem{qmga} A. T. Gabriel, T. Meyer and G. Germano 
{\it J. Chem. Theory Comput.} {\bf 4}, 468 (2008).
\bibitem{tweezers_nem} M. Yada {\em et al.} {\it Phys. Rev. Lett.} {\bf 92}, 185501 (2004); I. I. Smaluykh {\em et al.} {\it Phys. Rev. Lett.} {\bf 95}, 157801 (2005); I. I. Smalyukh {\em et al.} {\it Appl. Phys. Lett.} {\bf 86}, 021913 (2005).
\bibitem{tweezers_twist} U. Tkalec {\em et al.} {\it Phys. Rev. Lett.} {\bf 103}, 127801 (2009).
\bibitem{3Dmanipulation} These might include simultaneous 3D imaging and laser manipulation of defect clusters; see I. I. Smalyukh {\em et al.} {\it Optics Express} {\bf 15}, 4359 (2007).
\bibitem{hasimoto}H. Hasimoto {\em J. Fluid. Mech.} {\bf 5}, 317 (1959).


\end{thebibliography}
\end{document}